\documentclass{article}
\usepackage{amsmath}
\usepackage{amssymb}
\usepackage[final]{graphicx}
\begin{document}

\title{Growth and Optimality in Network Evolution}

\author{Markus Brede\\ CSIRO Centre for Complex System Science,\\ F C Pye Laboratory,
GPO Box 3023, Clunies Ross Street\\
Canberra ACT 2601, Australia\\
email:Markus.Brede@Csiro.au\\
phone: +61 62465628}
\maketitle

\begin{abstract}
In this paper we investigate networks whose evolution is governed by the interaction of a random assembly process and an optimization process. In the first process, new nodes are added one at a time and form connections to randomly selected old nodes. In between node additions, the network is rewired to minimize its pathlength. For timescales, at which neither the assembly nor the optimization processes are dominant, we find a rich variety of complex networks with power law tails in the degree distributions. These networks also exhibit non-trivial clustering, a hierarchical organization and interesting degree mixing patterns.\\

\noindent {\bf key words: complex networks, optimization, hierarchies, scale-free networks}
\end{abstract}


\section{Introduction}

Inspired by the finding that the topologies of a surprisingly large class of complex systems can be described by generic classes of networks, the last decade has seen a very rapid development in network modelling, cf. \cite{nets} for reviews. A particular emphasis has been put on two ubiquitous classes of networks: scale-free (SF) \cite{Barabasi} and small world (SW) \cite{Strogatz} networks. Many real-world networks display both features. SF networks are characterized by a degree distribution that has a power law tail of the form $P(k)\sim k^{-\alpha}$ with exponents $\alpha$ in the range between $2$ and $3$. SW networks combine short average pathlengths with a strong local cohesiveness expressed by high densities of triangles.

So far, most efforts to explain the formation SF networks have concentrated on assembly models using variants of preferential attachement. In these it is usually assumed that the network is assembled node by node over time, new nodes establishing connections to old nodes with attachment probabilities in proportion to the old nodes' degrees. The seminal studies which pioneered this approach are \cite{Barabasi,Dorogovtsev,Redner,Bianconi}.

Another line of thought to explain empirically observed network characteristics is via optimization models, assuming that the system under observation represents the end point of some optimization process. Examples to this approach include various models of link cost-pathlength optimization \cite{Sole,Gopal,MB-1,MB-2}, network models for optimal traffic \cite{Banavar}, linearly stable dynamics \cite{Variano,MB1}, and optimal synchronization \cite{Donetti,MB0}. All of these optimization models, however, consider systems that are static in size \cite{Remark}. For specific parameters SF networks have been obtained from an optimization model which minimized a trade-off between cost of links and average pathlengths \cite{Sole}. However, SF networks from optimization principles only seem to arise at transition points between different parameter regimes. Thus, to find SF networks from static optimization usually requires the fine tuning of some parameter -- clearly not a robust explanation.

In this paper we consider networks where both processes: assembly and optimization play a role. Naturally, the ratio of time scales of network assembly and optimization will be important to understand the evolution of such systems. Clearly, if assembly is slow in comparison to the optimization process, the topology of the evolved network can essentially be understood by a static optimization problem. Contrariwise, if network assembly is fast, the assembly mechanism will be the dominant process which shapes network topology. As we discussed above both of these limiting cases have been considered extensively in the literature before, in which it was always assumed that time scales for assembly and optimization are well separated and thus, that possible interactions between the two processes can be neglected. What, however, about the continuum of timescales in between, where both processes interact? This is the problem we will address in the present study.

\section{Model Description}

As an example for an optimization process to illustrate our point we consider the optimization for shortest average shortest pathlength,
\begin{align}
 d=\frac{2}{N(N-1)} \sum_{i<j} d(i,j),
\end{align}
where $d(i,j)$ is the length of the shortest path that connects the nodes $i$ and $j$ on the network. As an assembly process we consider the addition of nodes which form links to randomly selected nodes from the old network. For simplicity, we restrict the study to undirected binary links. The combination of an assembly and optimization mechanism appears of interest, since pathlength optimization typically generates star networks with one dominant hub node \cite{Sole,MB1}, while random node assembly generates networks with exponential degree distributions \cite{Barabasi}, i.e. networks without hub nodes. Thus two opposing tendencies are combined: a pressure towards hub formation resulting from the optimization process and a tendency against link accumulation, stemming from the node assembly. Below we will explore typical networks that are formed when the timescales of both processes are systematically varied.

More precisely, we consider a model of network assembly and evolution given by the following steps
\begin{itemize}
 \item[1.] Start with a small random network.
 \item[2.] Add a new node. The new node forms $k$ connections to randomly selected `old' nodes.
 \item[3.] Let the network evolve for optimal shortest pathlength. In more detail, randomly select a node $n$ and rewire one of its links to a randomly chosen new neighbor. If the average shortest pathlength of this node, i.e. $d_n=1/(N-1)\sum_{i\not= n} d(n,i)$ is reduced this way, the rewiring is accepted. If not, we proceed with the original network configuration. The optimization step 3. is repeated for $T$ times, if $T$ is not an integer or $T<1$ the fractional part of $T$ is interpreted as a probability that step 3. takes place an additional time. By normalizing the timescale of the assembly process to one unit of time, one can interpret the parameter $T$ as the ratio of the timescales of optimization and assembly.
 \item [4.] Proceed with 2. until the final network size has been reached.
\end{itemize}

\begin{figure}[tbp]
\begin{center}
\includegraphics [width=.65\textwidth]{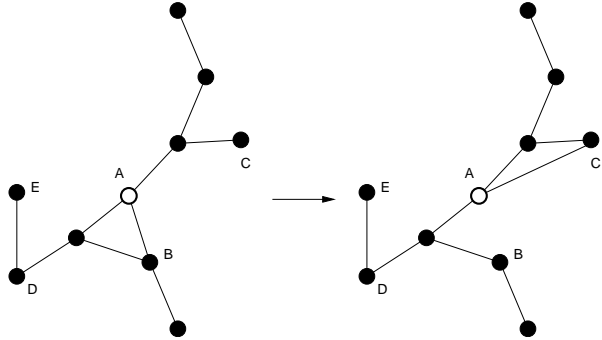}
 \caption{Illustration of the local rewiring. Node A is the target node and node $C$, which is $r=2$ steps away from A is chosen as the new destination node. Extending the requirement for strictly local rewiring at range $r=2$, new destination nodes can be at distance a distance $r>2$ from $A$. For instance, node $E$ would be a possibly choice for a destination node for $r=3$.  }
\label{FD}
\end{center}
\end{figure}

Later, instead of step 3. we also consider a variant of the optimization, where the search and rewiring is local. In this case we restrict new target nodes of the rewired link from $n$ to the neighborhood of $n$, i.e. nodes up to range $r=2$ away from $n$. An illustration of local rewiring schemes is given in figure \ref{FD}. The motivation for considering this local step is twofold. First, one may associate it with limited information of  agents that try to improve their position in the network, but only have local knowledge, i.e. knowledge about their neighbors. Second, one may also think about it as resulting from resource limitations when the nodes are embedded in space, such that agents can only form connections to nodes already close by.

Before proceeding, let us first briefly revisit the static problem, i.e. $T\gg 1$, treated, eg., in \cite{Sole,MB1}. Shortest pathlengths in a network with given number of links are realized in star-like networks. Depending on the number of links, a core with a dominant hub node, which links to all other nodes will form. Nodes other than the hub node have low degrees and, apart from the connection to the hub node, they have only a few connections to other such periphery nodes. Thus, pathlength-optimal networks are very heterogeneous networks, distinguished by the presence of a super hub node. It is worthwhile to note that the situation we treat here differs from \cite{Sole,MB1} in that nodes optimize their respective shortest pathlengths competitively, attempting to maximize individual fitnesses without consideration for the global good. In principle, in particular when link costs are explicitely considered, under a competitive scheme globally optimal configurations can become unstable \cite{MBS}. However, since the optimizing nodes only rewire their connections and do not eliminate links, numerical simulations indicate that the pathlength-optimal star-configuration is the only globally stable attractor of the dynamics.

As for the second limiting case $T\ll 1$, random assembly, it is known that growing networks where new nodes form connections to old nodes with equal probability leads to networks with exponentially decaying degree distributions \cite{Barabasi}. Thus, importantly, the attachment process does not lead to hub formation, SF degree distributions or SW behaviour.

\begin{figure}[tbp]
\begin{center}
\includegraphics [width=.95\textwidth]{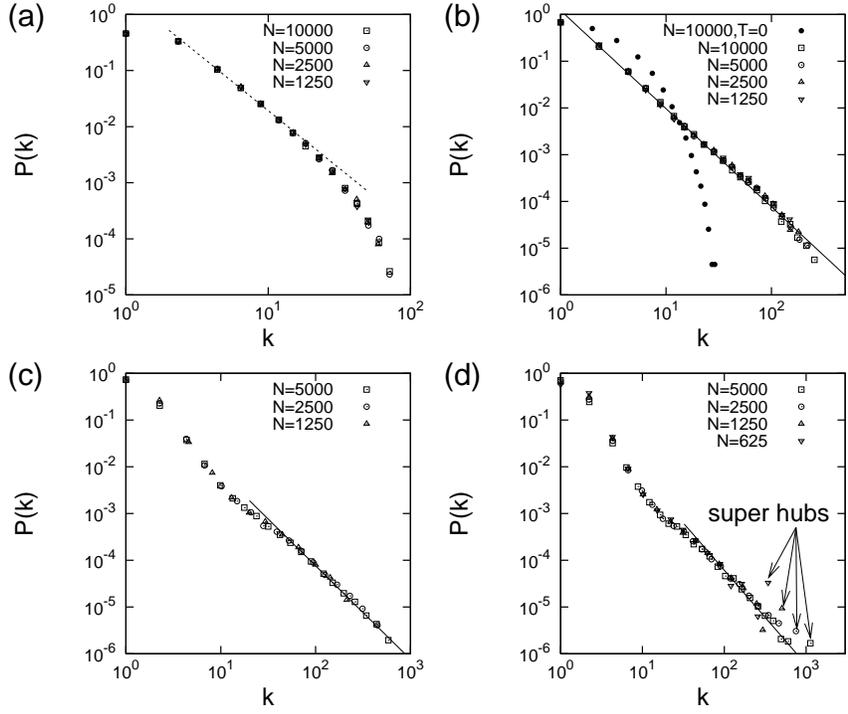} 
\caption{Degree distributions of networks with $\langle k\rangle=2$ evolved for (a) $T=10$, (b) $T=70$, (c) $T=350$ and (d) $T=1000$ for network sizes $N=625$, $N=1250$, $2500$, $5000$ and $10000$. The lines indicate power laws with exponent $\alpha=2$. Panel (b) additionally compares to a network of size $N=10000$ constructed for $T=0$ (no optimization). Averages are over at least 10 networks and data have been binned exponentially.}
\label{FA}
\end{center}
\end{figure}

\section{Global Optimization and Growth} 

In the following we numerically construct large networks of different sizes for systematically varied values of $T$. By displaying the degree distributions for some prototypical situations figure \ref{FA} summarizes some of the main results. Depending on the ratio between assembly and optimization timescales, four parameter regimes can be distinguished. In the first two regimes the degree distribution can be described by
\begin{align}
 P(k) \propto k^{-\alpha} F(k/k_N),
\end{align}
where $F(x)$ is a finite-size scaling function with $F(x)\approx \textrm{const.}$ for $x\ll 1$ and $F(x)\propto \exp(-x)$ for $x\gg 1$. The quantity $k_N$ gives a typical system size dependent cut-off degree at which the power law behaviour breaks down due to the finite system size $N$ and an exponential decay sets in. For SF networks one has $k_N\propto N^\delta$, with $\delta>0$ such that the power law regime in the degree distribution grows with increasing system size \cite{nets}.

The first regime in figure \ref{FA} corresponds to the case of small optimization times $T$, cf. panel (a) for $T=10$. One finds that the degree distributions are well described by a power law with exponent $\alpha=2$ for small degrees. However, the probabilities of larger degrees decay exponentially. The coincidence of the curves for system sizes of different orders of magnitude in panel (a) demonstrates that the cut-off length $k_N$ is independent of the system size, such that the resulting degree distributions have essentially exponentially decaying tails. 

With increasing $T$, typical cut-offs for exponential behaviour are shifted towards larger degrees, till eventually power law behaviour with exponent $\alpha=2$ governs all scales (and $k_N$ grows algebraically with the system size), cf. panel (b) for $T=70$. For larger $T$, low degree nodes start to lose connections disproportionately, while a fat tail of hub nodes extends to larger and larger degrees [panel (c) for $T=350$]. Finally, `super-hub' nodes appear, that are well-separated from the hub nodes in the power law tail, cf. panel (d) for $T=1000$.

\begin{figure}[tbp]
\begin{center}
\includegraphics [width=.65\textwidth]{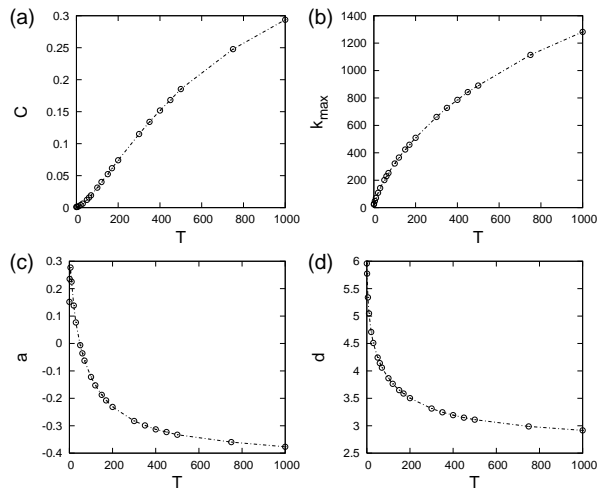} 
\caption{Dependence of network characteristics on the ratio of optimization and assembly time scales $T$: (a) clustering coefficient $C$, (b) max. degree $k_\text{max}$, (c) assortativeness $a$, and (d) average shortest pathlength $d$. The data have been collected for networks of size $N=5000$ and represent averages over at least 100 independent runs per value of $T$.} 
\label{FE}
\end{center}
\end{figure}

We proceed with an analysis of the networks in the different regimes. To characterize the networks, we analyse them in terms of the maximum degree $k_\text{max}=\max_i k_i$, average clustering coefficient \cite{Strogatz}, degree correlations as measured by the assortativeness $a$ defined in \cite{Newman}, and the average shortest pathlength $d$. Figure \ref{FE} gives the dependence of these key network statistics on $T$. Notably, increasing the timescales of the optimization process $T$ promotes hub formation [cf. Fig. \ref{FE}b], while also leading to more and more cliquish [cf. Fig. \ref{FE}a] and disassortative [cf. Fig. \ref{FE}c] network arrangements. Not surprisingly also, larger shares of optimization time compared to assembly time cause the networks to become smaller and smaller [cf. Fig. \ref{FE}d].

\begin{figure}[tbp]
\begin{center}
\includegraphics [width=.65\textwidth]{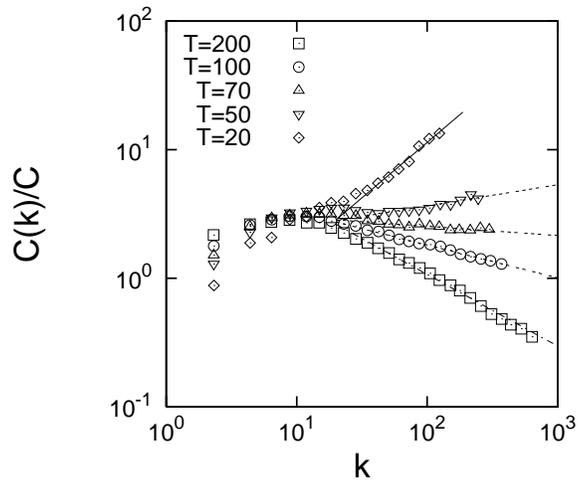} 
\caption{Dependence of the normalized clustering coefficient on degree for $N=10000$ and various values of $T$. Clustering coefficients are normalized by their averages $C$. Note the change at around $T=70$, when clustering starts to decay with degree.  Power-law behaviour $C(k)\sim k^{-\beta}$ (straight lines indicate best fits) for large $k$ points to a hierarchical organization: $\beta=0.56$ for $T=200$, $0.26$ for $T=100$, $0.07$ for $T=70$, $-0.16$ for $T=50$ and $-0.9$ for $T=20$.}
\label{FB}
\end{center}
\end{figure}

In many real-world systems, the cliquishness appears independent of system  size and the arrangement of cliques is hierarchical \cite{Ravasz}. We have conducted a scaling analysis of the dependence of the clustering coefficient on the degree and on the system size. The first is well described by a power law dependence 
\begin{align}
C(k)\sim k^{-\beta}
\end{align}
for large $k$, with exponents $\beta$ depending on $T$. Generally, for small $T$ positive $\beta$ is found, but the exponents decrease quickly to values around $\beta=0.5$ when $T$ is increased, thus indicating a hierarchical organization of cliques for large $T>100$, see figure \ref{FB}. The parameters around $T=70$ for which networks in which the power law dependence $P(k)\propto k^{-\alpha}$ with $\alpha \approx 2$ extends over all scales of the degree distribution [Fig. \ref{FA}b] have been generated, represent the boundary case between growth and decay in the $C(k)$-dependence. This indicates two different mechanisms to achieve small networks for small and large $T$. Essentially, for small $T$ no tail of hubs is formed. The network evolution generates a strongly cliquish core of nodes with larger than average degree. For large optimization times, the optimization procedure weeds out connections between hubs and hub competition leads to a less hierarchical organization of the hubs. This organization can be further classified by correlations between degrees of adjacent nodes.

\begin{figure}[tbp]
\begin{center}
\includegraphics [width=.65\textwidth]{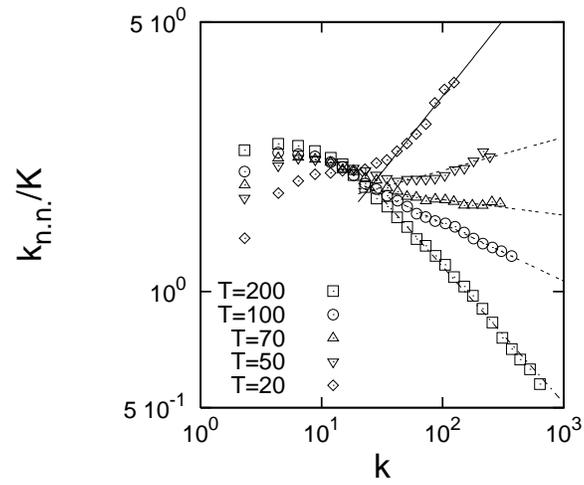} 
\caption{Dependence of the average nearest neighbor degree on the degree for $N=10000$ and various values of $T$. Average nearest neighbour degrees are normalized by their average $K$. Note the change in the dependence at $T=70$. The straight lines indicate power laws with exponents $\gamma=0.35$ for $T=200$, $0.15$ for $T=100$, $0.05$ for $T=70$, $-0.09$ for $T=50$ and $-0.4$ for $T=20$.}
\label{FC}
\end{center}
\end{figure}

Given the set of neighbors ${\cal N}(i)$ of a node $i$, let us define the average neighbor degree of $i$ via 
\begin{align}
k_\text{nn,i}=1/k_i\sum_{j \in {\cal N}(i)} k_j
\end{align}
and distinguish classes of nodes by degree. Following \cite{Pastor} we then introduce a nearest neighbor degree function $k_\text{nn} (k)$, that gives the average over all nodes with degree $k$. Simulation data for this function reveal that $k_\text{nn} (k)$ generally follows a power law 
\begin{align}
k_\text{nn}(k)\propto k^{-\gamma}
\end{align}
 for large $k$. As already expected from the dependence of the aggregate assortativeness on $T$ (cf. figure \ref{FE}c), a strong dependence of nearest neighbor correlations on $T$ is notable.  For small $T$ one has $\gamma<0$ and observes a transition towards $\gamma>0$ as $T$ is increased, cf. figure \ref{FC}. The range of parameters around $T\approx 70$ again marks the boundary at which $\gamma$   changes its sign. Hence, for $T<70$ networks form a hierarchy of large degree nodes that are preferentially linked to each other. Forming links to hub nodes of slightly smaller degrees the hub node with the largest degree is located in the centre. Hub nodes of this next level form links to nodes of slightly smaller degree while low degree nodes are mostly linked to each other and are located at the periphery of the network. This contrasts with typical network structures generated for large $T$, when the optimization process weeds out redundant connections between the hub nodes. In this case, central hubs preferentially link to low degree nodes, entailing a strongly disassortative arrangement.

Similar to the preferential attachment model \cite{Barabasi} an analysis of the scaling with system size manifests that the clustering eventually decays to zero in the limit of very large systems (data not shown).

In sum, the interaction of assembly and global optimization processes at various timescales can generate very skewed network topologies. A transition timescale ratio $T\approx 70$ exists at which degree distributions follow a power law with exponent $\alpha=2$. Similar to the model of Ref. \cite{Sole}, explaining power law degree distributions by the above mechanism would require the fine-tuning of the timescale ratio to $T\approx 70$. Below, we will extend the model towards local rewiring procedures. As we will see, in this case a broad range of timescale ratios for which degree distribitions follow power laws exists such that no fine-tuning of parameters to obtain power-laws is required. 

\section{Local Optimization and Growth} 

The above finding motivates us to also consider a different optimization process, where rewirings are restricted to local connections (cf. figure \ref{FD} above). Note, that in such a local rewiring process the number of triangles can only increase (if the rewired connection was not part of a triangle) or stay the same (if the rewired connection was part of a triangle). Nevertheless, since the presence of communities in sparse networks does not promote short pathlengths, clustering coefficients stay well below their maximum values even for relatively large $T$. 

It is also worthwhile pointing out that the rewiring process without selection does not lead to very skewed or SF degree distributions. The reason is that even though nodes attract new connections in proportion to their degree, connections are also chosen for rewiring in proportion to a node's degree. Both effects compensate such that the resulting degree distributions remain narrow. A further immediate observation is about characteristic timescales of network optimization subject to local or global rewiring. Essentially because in local rewiring new destination nodes are suggested in proportion to the destination nodes' degree  the likelihood of accepting a rewiring suggestion is enhanced. For this reason the local optimization is much more effective than the global process and characteristic timescales of the local optimization are much shorter than for the global process.

\begin{figure}[tbp]
\begin{center}
\includegraphics [width=.95\textwidth]{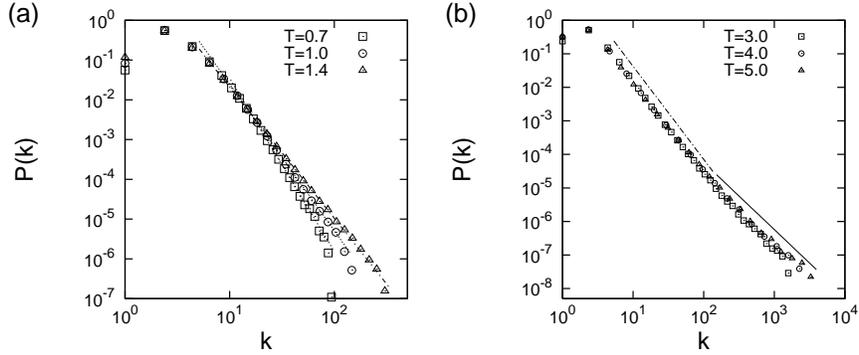}

\caption{Degree distributions for networks of size $N=50000$ constructed by the local optimization scheme with range $r=2$ for (a) $T=0.7$, $1.0$ and $1.4$ and (b) $T=3.0$, $4.0$ and $5.0$. In panel (a) the lines indicate power laws with exponents $\alpha=4.35$, $3.69$ and $3.3$, in (b) they indicate power laws with $\alpha=2.8$ for small degrees and $\alpha=2.0$ for the tail of large degrees. At around $T=5.0$ `super-hubs' start to emerge. Averages over 30 networks, distributions are binned logarithmically.}
\label{FF}
\end{center}
\end{figure}

With the data for degree distributions of networks assembled subject to local rewiring in figure \ref{FF} we proceed with an analysis of the local optimization scheme for range $r=2$. Apart from different characteristic timescales, results for the degree distributions are analogous to the case of globally optimized networks. Very short optimization timescales do not allow for the formation of networks with very skewed degree distributions. However, from around $T=0.7$ onwards, degree distributions start to exhibit power law tails. Increasing the optimization timescale, the exponent characterizing the power law tails declines. However, as panel (a) of figure \ref{FF} demonstrates the distributions still clearly exhibit power-laws over around two orders of magnitude. For example, we find power laws with exponents $\alpha=4.35$, $3.69$ and $3.3$ for $T=0.7, 1.0$ and $1.4$. At around $T=1.5$ a tail of hub nodes that deviates from the power law behaviour for nodes of small degree develops, cf. panel (b). For $T\leq 5$ this tail is well-described by a power law with exponent $\alpha=2$. At around $T\approx 5$ `super-hubs' emerge and gradually attract more and more connections when further increasing $T$. 

A diagram of further average network characteristics such as clustering coefficients, maximum degrees, assortativeness or pathlengths also displays very similar characteristics to the one for the global optimization scheme displayed in figure \ref{FE}, but we don't discuss them in detail here.

It is of interest to note that, even though power law tails emerge for a broad range of timescale ratios, local rewiring only allows for the assembly of networks with power-law degree distributions with exponents $\alpha>3$. However, most real-world networks in biological and other applications have been found to have exponents in the range $2<\alpha<3$. Hence, optimization and strictly local rewiring with range $r=2$ cannot provide explanation for the structure of real-world networks. However, as we will see below, relaxing the constraint of strictly local rewiring, a much larger configuration space of possible network topologies can be explored.

\section{Between local and global} 

Above, we considered a local rewiring scheme where rewirings were always directed towards nodes at distance $r=2$ from the target node. One can extend this concept by gradually extending the set of new destination nodes to nodes not farther away than an arbitrary range $r\geq 2$ \cite{Remark1} (and refer to figure \ref{FD} for an illustration). Since $r\to \infty$ recovers the case of global optimization, tuning $r$ allows to interpolate between the local and global optimization schemes. Let us assume a fixed timescale ratio $T$ between network assembly and optimization. From the analysis of the limiting cases of global and local rewiring in the previous sections, one can understand the implications of a rewiring parameter $r>2$ on the evolved network topologies. Increasing $r$ to values that are close to the characteristic pathlengths of the network reduces the ``efficiency'' of the optimization and dominant hub nodes are less likely to appear. Because of this a scenario for a choice of $r_1$, $T_1$, corresponds to a scenario $r_2<r_1$ with $T_2<T_1$. Hence, one expects that an increase in $r$ allows for the assembly of networks with less skewed degree distributions. Trivially, relaxing the constraint for local rewiring will also reduce cliquishness.
 
\begin{figure}[tbp]
\begin{center}
\includegraphics [width=.95\textwidth]{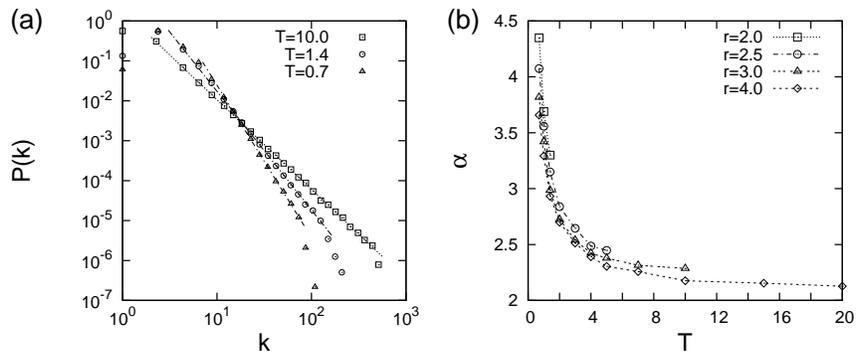}
\caption{(a) Degree distributions of networks generated for $N=50000$, $r=3$ and $T=0.7$, $1.4$ and $10.0$. The lines indicate power laws with exponents $\alpha=3.8$, $3.0$ and $2.28$, respectively.  (b) Dependence of the exponents $\alpha$ on the optimization timescale $T$ for different ranges $r=2$ (squares), $2.5$ (spheres), $r=3$ (triangles), and $r=4$ (diamonds). The values have been determined from best fits to the logarithmically binned tails of the degree distributions constructed from at least 10 networks of size $N=50.000$. For $r=2$, $2.5$, $3$ and $4$ at approximately $T=1.5$, $5$, $10$ and $20$ the power laws describing small and large degrees start to diverge, cf. text. } 
\label{FH}
\end{center}
\end{figure}

Interestingly, similar to what we observed for $T<1.5$ for $r=2$ above, for different values of $r$ networks with different steepness of the decay of the degree distribution can be constructed. However, unlike for $r=2$, for $r>2$ degree exponents in the range $2<\alpha<3$ can be found. This is demonstrated by the simulation data displayed in figure \ref{FH}a, in which we compare some example degree distributions for networks generated with $r=3$ and different $T$. A more comprehensive analysis of degree distributions that can be obtained for various choices of $r$ and $T$ is given in figure \ref{FH}b. Two observations are in order. First, increasing $r$ beyond strictly local rewiring for $r=2$, degree distributions with exponents $\alpha$ in the range $2.1<\alpha<4.5$ can be obtained.  The larger $r$ the smaller the exponent characterizing the degree distribution. Second, the power law regime is again limited to a range between very low and large values of $T$, outside of which either the assembly process dominates or when hub formation sets in. It is difficult to give strict bounds for the range of $T$ values for which power-law behavior is found, but the interval covers about one order of magnitude for rewiring between global and strictly local rewiring (e.g. from about $T=0.5$ to $T=20$ for $r=4$). Clearly, no fine tuning of parameters is required to find scale-free networks.

\begin{figure}[tbp]
\begin{center}
\includegraphics [width=.65\textwidth]{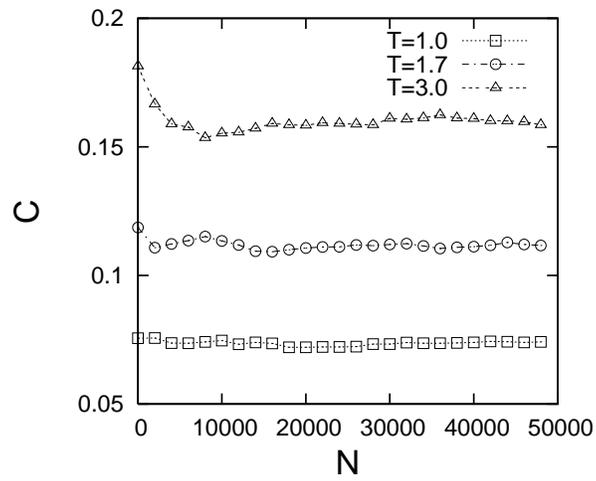} 
 
\caption{Scaling analysis of the average clustering coefficient $C$ with system size for local rewiring with $r=3$. Note, that $C$ converges to a constant value in the limit of large system sizes. }
\label{FD1}
\end{center}
\end{figure}

Another important point is, that also for $2\leq r\leq 3$ strongly cliquish networks are formed. This is illustrated in figure \ref{FD1}, which gives some examples of the change in cliquishness with growing system sizes for $r=3$ and different $T$. It is important to note that unlike the case of the dependence of clustering on the system size for global optimization, for every $T>0$ clustering coefficients converge towards constant non-trivial values for large systems.

\section{Summary and Conclusions}

In summary, in this paper we have analyzed a class of networks that emerge from the interplay of a network assembly process and an optimization process. A systematic investigation of the parameter space reveals that the ratio of timescales $T$ between both processes is an important determinant of network structure. For low $T$, network assembly dominates and the constructed networks have narrow degree distributions. In the opposite case, for large $T$, network optimization dominates, leading to the formation of networks with bimodal degree distributions. These networks are characterized by a periphery of low degree nodes and a core of a few ``super-hub'' nodes. For $T\to \infty$ they evolve into star networks.

Interestingly, we have found a parameter range ``in between'', in which the structural influences of network  assembly and network optimization are balanced. In this parameter regime we find network structures exhibiting power law degree distributions with exponent $2$ which are further classified by a hierarchical organization, non-trivial degree mixing and very short pathlengths.

Additional to its characteristic timescale $T$, the network optimization process can be classified by a second parameter, the locality $r$ of the rewiring mechanism. The parameter can be understood as a measure for the information about the system that is available to individual nodes. Varying $r$ we have explored situations ranging from strictly local rewiring (each node has only information about and access to its immediate neighbors) to global rewiring (every node has global information and access). Our analysis points out that this mechanism can generate a very rich variety of SF complex networks with various power law exponents in the range $2.1<\alpha<4.5$, all of which also exhibit non-trivial clustering and interesting patterns of degree mixing. Importantly, the mechanism leads to the emergence of power laws for a broad parameter range, such that no fine-tuning of parameters is required to explain SF behaviour from network optimization.

\bibliographystyle{alj}

\begin{thebibliography}{}

\bibitem{Remark}
A notable exception to this are the studies \cite{Gleiser} which consider growing networks, where new nodes attach and adjust their links to optimize a measure of synchronization. Distinct from \cite{Gleiser}, however, we consider optimization as a process affecting the whole network and not only recently added nodes.

\bibitem{Remark1}
Non-integer values are interpreted as averages, i.e. for non-integer value $r$ in an optimization step new connections are made to nodes not farther away than distance $\lfloor r\rfloor +1$ with prob. $q=r-\lfloor r\rfloor$ and, alternatively, to nodes not farther away than distance $\lfloor r\rfloor$ with prob. $1-q$.

\bibitem{nets}	
         Albert, R., \& Barab\'asi, A.-L. (2002). Statistical mechanics of complex networks. {\it Reviews of Modern Physics}, 74, 47--97.\\
         Newman, M. J. E. (2003). The structure and function of complex networks, {\it SIAM Review}. 45, 167--256.\\
         Boccaletti, S., Latora, V., Moreno, Y., Chavez, M., \& Hwang, D.-U. (2006). Complex networks: Structure and dynamics. {\it Physics Reports}, 424, 175--308.

\bibitem{Barabasi}
 Albert, R., \& Barab\'asi, A.-L. (1999). Emergence of scaling in random networks. {\it Science}, 286, 509--512.

\bibitem{Bianconi}
 Bianconi, G., \& Barab\'asi, A.-L. (2001). Competition and multiscaling in evolving networks. {\it Europhysics Letters}, 54, 436--442.

\bibitem{MB0}	
         Brede, M. (2008). Construction principles for highly synchronizable sparse directed networks. {\it Physics Letters A}, 372, 5305--5308.\\
	 Brede, M. (2008). Synchrony-optimized networks of non-identical Kuramoto oscillators. {\it  Physics Letters A}, 372, 2618--2622.\\	
         Brede, M. (2008).  Locals vs. global synchronization in networks of non-identical Kuramoto oscillators. {\it European Physical Journal B}, 62, 87--94.

\bibitem{MB1}	
         Brede, M., \& de Vries, B. J. M., Networks that optimize a trade-off between efficiency and dynamical resilience. {\it Physics Letters A}, 373, 3910--3914.

\bibitem{MB-1}
 Brede, M. (2010). Small worlds in space: Synchronization, spatial and relational modularity. {\it Europhysics Letters}, 90, 60005.

\bibitem{MB-2}
 Brede, M. (2010). Optimal synchronization in space. {\it Physical Review E}, 81, 025202.

\bibitem{MBS}	
         Brede, M. (2010). Coordinated and uncoordinated optimization of networks. {\it Physical Review E}, 81, 066104.

\bibitem{Banavar}
Colizza, V., Banavar, J. R., Maritan, A., \& Rinaldo, A. (2004). Network structures from selection principles. {\it Physical Review Letters}, 92, 198701.

\bibitem{Donetti}	
  Donetti, L., Hurtado, P. I., \& Mu$\tilde{\text{n}}$oz, M. A. (2005). Entangled networks, synchronization, and optimal network topology, {\it Physical Review Letters}. 95, 188701.

\bibitem{Dorogovtsev}
 Dorogovtsev, S. N., \& Mendes, J. F. F., \& Samukhin, A. N. (2000). Structure of growing networks with preferential linking. {\it Physical Review Letters}, 85, 4633.

\bibitem{Sole}
Ferrer i Cancho, R., \& Sol\'e, R. V. (2003). Optimization in complex networks. In R. Pastor-Satorras, M. Rubi, \& A. Diaz-Guilera (Eds.), {\it Statistical mechanics of complex networks} (pp. 114--126). Berlin: Springer.

\bibitem{Gleiser}	
         Gomez Portillo, I. J., \& Gleiser, P. M. (2009). An adaptive complex network model for brain functional networks. {\it PLoS ONE}, 4, e6863.

\bibitem{Gopal}
 Mathias, N., \& Gopal, V. (2001). Small worlds: How and why. {\it Physical Review E}, 63, 021117.

\bibitem{Redner}
 Krapivsky, P. L., \& Redner, S., \& Leyvraz, F. (2000). Connectivity of growing random networks. {\it Physical Review Letters}, 85, 4629--4632.


\bibitem{Newman}	
        Newman, M. E. J. (2002). Assortative mixing in networks. {\it Physical Review Letters}, 85, 208701.

\bibitem{Pastor}	
        Pastor-Satorras, R., V\'asquez, A., \& Vespignani, A. (2001). Dynamical and correlation properties of the internet. {\it Physical Review Letters}, 87, 258701.

\bibitem{Ravasz}	
        Ravasz, E., Somera, A. L., Mongru, D. A., Oltvai, Z. N., \& Barab\'asi, A.-L. (2002). Hierarchical organization of modularity in metabolic networks. {\it Science}, 297, 1551--1555.

\bibitem{Variano}	
         Variano, E. A., McCoy, J. H., Lipson, H. (2004). Networks, dynamics, and modularity. {\it Physical Review Letters}, 92, 188701.

\bibitem{Strogatz}
 Watts, D. J., \& Strogatz, S. H. (1998). Collective dynamics of `small-world' networks. {\it Nature}, 393, 440--442.

\end{thebibliography}

\end{document}